\documentclass[twocolumn,reprint,showpacs,amsmath,amssymb,xcolor=dvipsnames,prl,superscriptaddress]{revtex4-1}
\pdfoutput=1

\usepackage[pdftex]{graphicx}
\usepackage{bm}
\usepackage[pdftex,colorlinks=true,citecolor=blue,urlcolor=blue,linkcolor=black]{hyperref}
\usepackage{braket}
\usepackage[dvipsnames]{xcolor}
\usepackage[version-1-compatibility,binary-units,input-product=x,abbreviations=true]{siunitx}

\begin{document}


\title{Observation of Pauli Crystals}
\author{Marvin~Holten}
    \email{mholten@physi.uni-heidelberg.de}
	\affiliation{Physikalisches Institut der Universit\"at Heidelberg, Im Neuenheimer Feld 226, 69120 Heidelberg, Germany}
\author{Luca~Bayha}
    \email{bayha@physi.uni-heidelberg.de}
	\affiliation{Physikalisches Institut der Universit\"at Heidelberg, Im Neuenheimer Feld 226, 69120 Heidelberg, Germany}
\author{Keerthan~Subramanian}
	\affiliation{Physikalisches Institut der Universit\"at Heidelberg, Im Neuenheimer Feld 226, 69120 Heidelberg, Germany}
\author{Carl~Heintze}
	\affiliation{Physikalisches Institut der Universit\"at Heidelberg, Im Neuenheimer Feld 226, 69120 Heidelberg, Germany}
\author{Philipp~M.~Preiss}
	\affiliation{Physikalisches Institut der Universit\"at Heidelberg, Im Neuenheimer Feld 226, 69120 Heidelberg, Germany}
\author{Selim~Jochim}
	\affiliation{Physikalisches Institut der Universit\"at Heidelberg, Im Neuenheimer Feld 226, 69120 Heidelberg, Germany}

\date{\today}

\begin{abstract}

The Pauli exclusion principle is a fundamental law underpinning the structure of matter. Due to their anti-symmetric wave function, no two fermions can occupy the same quantum state. Here, we report on the direct observation of the Pauli principle in a continuous system of up to six particles in the ground state of a two-dimensional harmonic oscillator. To this end, we sample the full many-body wavefunction by applying a single atom resolved imaging scheme in momentum space. We find so-called Pauli crystals as a manifestation of higher order correlations. In contrast to true crystalline phases, these unique high-order density correlations emerge even without any interactions present. Our work lays the foundation for future studies of correlations in strongly interacting systems of many fermions.

\end{abstract}
\maketitle


Correlated fermions lie at the heart of many open questions concerning quantum matter that remain unresolved to this day. 
Knowledge of the type and origin of correlations, especially of higher orders, is an essential cornerstone in the endeavor of solving these complex many-body systems \cite{Altman2004,Schweigler2017,Hodgman2017}. Strong correlations are, however, not exclusive to interacting particles as was demonstrated already in 1956 by the famous experiment of Hanbury Brown and Twiss \cite{Brown1956}. Bosons tend to occupy the same quantum state while multiple occupation of a single state is forbidden for fermions due to the Pauli exclusion principle. 

In a degenerate gas of neutral fermions, Pauli exclusion reveals itself through suppression of collisions \cite{DeMarco1999,DeMarco2001} and an effective Fermi pressure \cite{Truscott2001}. Antibunching has been observed directly in time-of-flight measurements \cite{Rom2006,Jeltes2007} or via the suppression of density fluctuations \cite{Mueller2010,Sanner2010}. Quantum gas microscopy advances have led to the first single atom resolved observation of Pauli blocking in the band insulating regime of a lattice potential \cite{Omran2015}.

Here, we extend single atom resolved measurements of fermionic correlations to continuous systems. We study ultracold, fermionic atoms that are confined to a two-dimensional harmonic oscillator potential. Even in the absence of interactions, the Pauli exclusion leads to local high-order density correlations between the atoms that go beyond a simple Fermi hole. The geometric patterns that we observe are clearly distinct from those in interaction driven systems and they have been termed Pauli crystals \cite{Gajda2016}. 

Pauli crystals only emerge at very low temperatures where the particles become quantum degenerate and their Fermi energy dominates over temperature and trap imperfections. This requires charge-neutral non-interacting systems and control on very low absolute energy scales \cite{Rakshit2017}. The structures act as a starting point for the study of correlations in continuous systems with single atom resolution.

\begin{figure}
    \centering
	\includegraphics{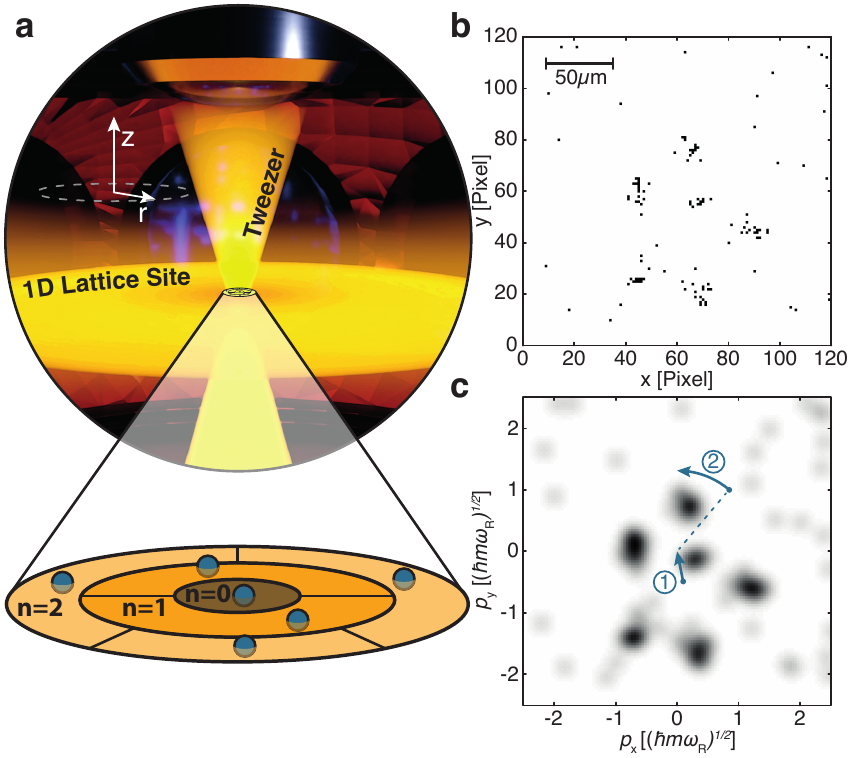}
    \caption{ \textbf{Sketch of the experimental setup.} The atoms are trapped in a single site of an attractive one-dimensional optical lattice in the vertical direction superimposed with a tightly focused optical tweezer (a, top). The degeneracy of the effectively two-dimensional harmonic confinement leads to the formation of a non-trivial shell structure (a, bottom). Binarized image of the $N=6$ closed-shell system taken with a single photon-counting EMCCD camera after a time-of-flight expansion (b). We extract the atom momenta by searching for local maxima in the lowpass-filtered image (c). All momenta are plotted in natural units of the harmonic oscillator confinement. To reveal correlations between the particles we subtract the center of mass motion (c, arrow 1) and rotate to a common symmetry axis (c, arrow 2).}
    \label{fig:main1}
\end{figure}

\begin{figure*}
    \centering
	\includegraphics{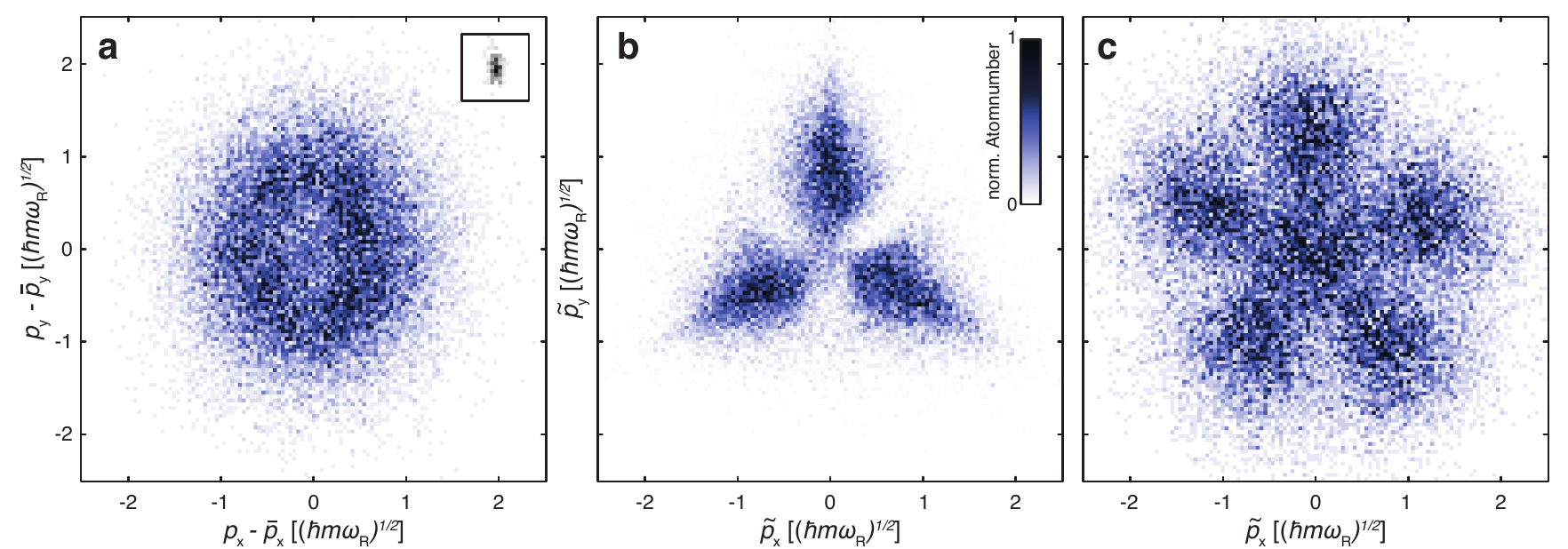}
    \caption{ \textbf{Pauli crystal measurements with $\bm{N=3}$ and $\bm{N=6}$ particles.} A 2D histogram of the measured momenta $p_\text{i}$ minus the center of mass momentum $\bar{p}$ leads to the one-particle relative momentum distribution of the $N=3$ system (a). The reduced density in the center is a result of the exact form of the harmonic oscillator eigenfunctions and not a consequence of anti-symmetrization. The inset shows a measurement of the momentum point spread function of our imaging setup plotted on the same scale. The strong correlations between the fermions only reveal themselves in the configuration probability densities for $N=3$ (b) and $N=6$ (c), where each experimental run has been rotated separately to a common symmetry axis.}
    \label{fig:main2}
\end{figure*}

The experimental observation of Pauli crystals relies on two essential capabilities: the preparation of $N$ non-interacting fermions in a well defined quantum state and the detection of $N$-body correlations in the relative positions or momenta of these particles. We perform our experiments with a balanced mixture of two hyperfine states of $^6 \text{Li}$ confined by the superposition of an optical tweezer and a single site of a one-dimensional optical lattice in the vertical direction (see Figure \ref{fig:main1} a). The large ratio between axial ($\omega_z = 2\pi \times 6560(6) \,\text{Hz}$) and radial ($\omega_\text{r} = 2 \pi \times 983(5) \,\text{Hz}$) trap frequencies allows us to work in a quasi-2D regime for sufficiently small temperature $T$ and particle number $N$. In this limit all the atoms occupy the motional ground state in the axial direction and the dynamics are limited to the harmonic confinement in radial direction.

The $n^\text{th}$ energy level of a symmetric two-dimensional harmonic oscillator is ($n$+1)-fold degenerate, leading to three lowest closed-shell configurations filled with 1, 3 and 6 fermions per spin state respectively (see Figure \ref{fig:main1} a). A spilling technique that was initially pioneered for one-dimensional systems \cite{Serwane2011a} and that we extended to two dimensions \cite{Bayha2020} allows us to reach these configurations filled with two spin components. The preparation fidelities are $92(7)\,\%$ for the 3+3 ($N_\uparrow+N_\downarrow)$ and $56(3)\,\%$ for the 6+6 ground states respectively. We work with a two component mixture instead of a single component gas since attractive interactions during the spilling sequence improve the preparation fidelity.
To create a non-interacting mixture, we make use of a Feshbach resonance \cite{Zuern2013} and adiabatically ramp the magnetic offset field $B$ to a zero crossing of the scattering length $a_\text{3D}$ at $568\,\text{G}$ once the system is initialized in the ground state. For all the measurements presented in the following, the atoms of only one of the two spin components are imaged.

We extract momentum correlations from our system by first mapping the initial momenta of the particles onto their position by a TOF expansion for a quarter trap period in a single lattice site with $\omega_\text{TOF} = 2\pi \times 20.7(5)\,\text{Hz}$ \cite{Murthy2014}. This is followed by the single atom detection fluorescence imaging scheme discussed in detail in Ref.\ \cite{Bergschneider2018}. By collecting on average 20 photons per atom on a single photon counting camera, this method allows us to detect atoms of a single spin component in free space with fidelities on the order of $95\,\%$ (see Figure \ref{fig:main1} b). Each image obtained in this way represents a single sample $p_\text{i,x}$ and $p_\text{i,y}$ of the in-situ momentum distribution for every particle $\text{i}=1...N$ of one of the spin components (see Figure \ref{fig:main1} c).

The harmonic confinement plays an important role for our measurements. We prepare our atoms in eigenstates of the harmonic oscillator. The real space wavefunctions are therefore given by Slater determinants that contain superpositions of Hermite polynomials \cite{som}. These wavefunctions are invariant under continuous Fourier transforms and therefore also invariant under our TOF expansion. The expansion simply corresponds to a magnification of the in-situ wave function by a factor of 50. This leads to an effective imaging resolution of approximately $200\,\text{nm}$. The natural scale of the harmonic oscillator states is given by $p_\text{0}=\sqrt{\hbar m \omega_\text{r}}$ or $l_\text{0}=\hbar/ p_\text{0} = 1.31(1) \,\mu\text{m}$.

\begin{figure}[hbt!]
    \centering
	\includegraphics{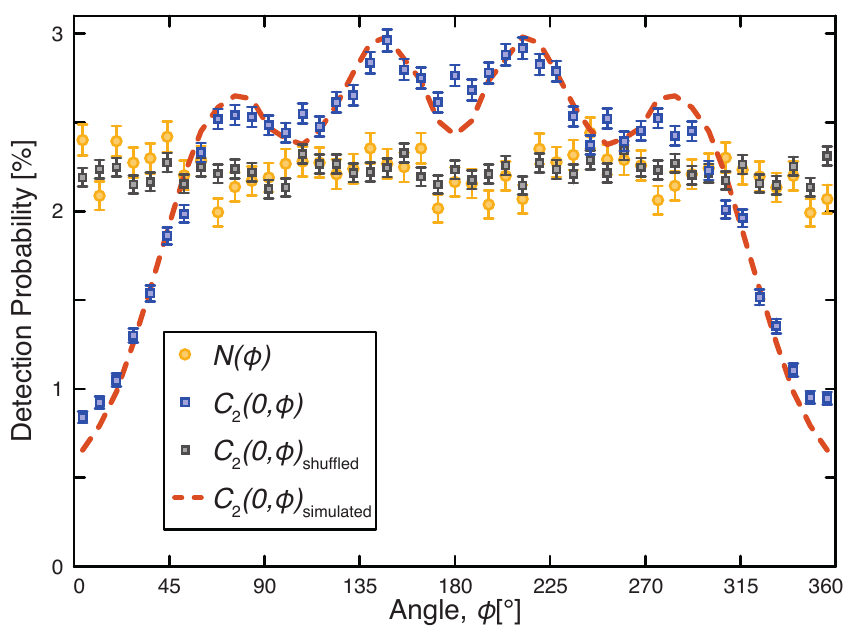}
    \caption{\textbf{Angle correlations for $\bm{N=6}$ particles.} The angular correlation function $C_2$ is computed after removing the center of mass motion by fixing one particle at $\phi_1=0$ and creating a histogram of the measured angles of the other particles. $C_2(0,\phi)$ shows 4 maxima in addition to the hole around $\phi=0$ as expected for the 5-fold symmetry of the Pauli crystal (blue points). The measurement agrees well with a Monte Carlo simulation (dashed line). Both the total angular atom distribution $N(\phi)$ (yellow points) and the correlation function of shuffled data (grey points) show no structure.}
    \label{fig:main3}
\end{figure}

Our measurements have been performed by preparing the system at the lowest accessible temperatures with two ($N=3$) or three ($N=6$) harmonic oscillator shells filled and collecting $9994$ and $19291$ TOF images respectively. Only images where the correct number of atoms have been detected are investigated further, leading to post selection rates of $25\,\%$ and $28\,\%$. We process these measurements as suggested by Ref.\ \cite{Gajda2016} to reveal correlations between the fermionic particles (see Figure \ref{fig:main1} c). In the first step, we subtract the respective center of mass momentum $\bar{p}$ from each set of momenta $p_\text{i}$. We find that the width of the center of mass momentum scales with the inverse square root of the total mass (i.e. $m_\text{tot}=N\cdot m$, where $m$ is the mass of one $^6$Li atom) as expected \cite{som}.

A histogram of the remaining relative momenta yields the one-particle momentum distribution (see Figure \ref{fig:main2} a). The latter expresses the probability to find one particle with momentum $p_\text{x}-\bar{p}_\text{x}$ and $p_\text{y}-\bar{p}_\text{y}$ when integrating over all possible momenta of all other particles. We stress that the one-particle momentum density can also be obtained from average density images without single particle resolution and does not reveal any higher order correlations. Our data, however, contains more information: we know the full configuration of all particles in every single realization of the experiment.

To extract correlations, a second processing step is necessary. Due to the radial symmetry of our system, the angle distribution of all particles $N(\phi)$ is homogeneous (see Figure \ref{fig:main3}). The rotational symmetry is only broken in each experimental realization by the measurement itself and the particles align with respect to a random axis that is different for each set of momenta $p_\text{i}$. We rotate each set independently to a new coordinate system $\tilde{p}_\text{i}$ by an angle that minimizes the distance to a chosen target configuration \cite{Gajda2016,som}.
Strong correlations in both the $N=3$ and the $N=6$ particle state become apparent immediately in the $N$ particle momentum configuration distributions that we obtain in this way (see Figure \ref{fig:main2} b,c).

The observation of these so-called Pauli crystals confirms theoretical predictions \cite{Gajda2016}. To rule out an artifact of our analysis, we compared to images that were obtained after shuffling the atom momenta between different experimental runs \cite{som}. This is especially important since it has been shown that other distance measures can cause a bias towards the target configuration \cite{Fremling2020}. In addition, we calculate the angular density-density correlation function $C_2(0,\phi)$ that expresses the probability of finding a second particle at an angle $\phi_2=\phi$, when one particle is fixed at $\phi_1=0$. The result for $N=6$ is shown in Figure \ref{fig:main3} and clearly shows the presence of four peaks, as expected for the Pauli crystal and in agreement with a Monte Carlo simulation \cite{som}.

\begin{figure}
    \centering
	\includegraphics{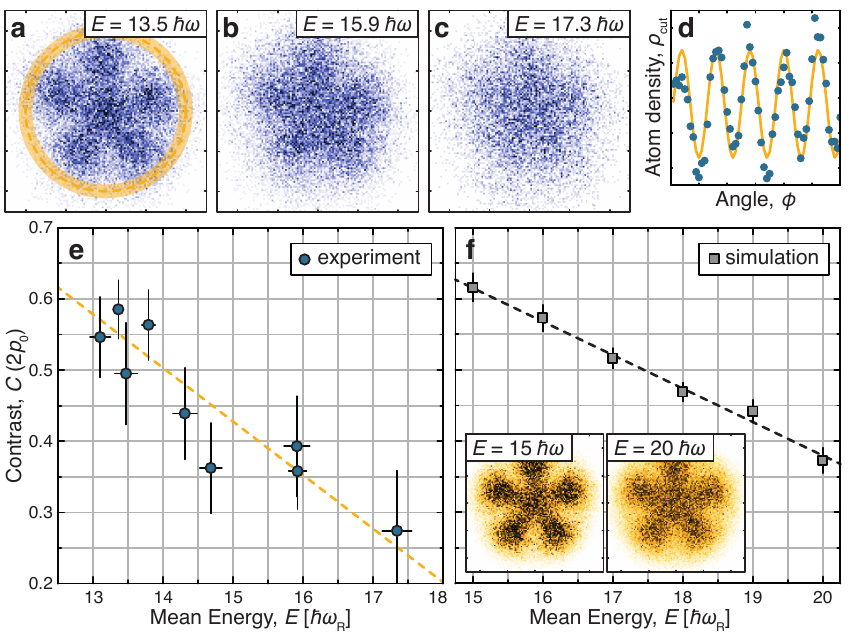}
    \caption{ \textbf{Melting the Pauli crystal.} (a-c) $N=6$ Momentum configuration distributions for different initial state energies. We find that the Pauli crystal melts quickly when we add energy to the initial state by modulating the trap potential. (d) We extract the contrast $C$ by fitting a sine function to the distribution at a fixed momentum $p=2p_\text{0}$ (highlighted by the yellow line in a). (e) The contrast reduces with increasing total mean energy of the system. (f) We compare our measurement to a Monte Carlo simulation, where we sample states from a thermal $N$-body density matrix. We follow the simulation procedure discussed in Ref. \cite{Rakshit2017} and include only states up to a maximal excitation energy of $6\hbar\omega_\text{r}$ to reduce the computational cost. The dashed lines are linear fits to the data.}
    \label{fig:main4}
\end{figure}

We find that the observed configuration distributions exhibit weaker modulation than the simulations we perform for systems at zero temperature. We quantify this effect through the contrast $C(p)$, which we define as one minus the ratio of the minima and maxima of a fit to the configuration distribution at a fixed radial momentum $p = 2p_\text{0}$ (see Figure \ref{fig:main4} a,d). Apart from technical limitations like the point spread function of our imaging setup or fluctuations of trap potentials, we identify the finite temperature of our initial state as the main cause of this reduction.

To study the effect of finite temperature in more detail, we ``melt" the $N=6$ crystal by increasing the mean energy of the initial state (see Figure \ref{fig:main4} a-c). To this end, we modulate the confining potential at twice the trap frequency $2\omega_\text{r}$ with variable amplitudes and take around 3000 images at each setting. Trap imperfections like anharmonicity and anisotropy in combination with small potential drifts cause the system to dephase on timescales much faster than the modulation time of $t=50\,\text{ms}$ and the excitation is therefore not coherent.

The total energy of the system is extracted from the momentum measurements and averaged over each data set. We find that the energy of the lowest temperature initial state, without any applied modulation, is $E = 13.1 \hbar \omega_\text{r}$. This value is about $5\,\%$ below the expected $N=6$ ground state energy of $E_\text{g} = 14 \hbar \omega_\text{r}$. Our measurement of the energy entails systematic uncertainties like the error on the frequency measurement of the expansion potential $\omega_\text{TOF}$ ($\approx 2\,\%$) or the magnification of our imaging setup that both enter quadratically. Taken together, these uncertainties may account for the observed shift to lower energies, which is systematic for all data points.

A comparison of the relative change in energy and contrast clearly displays the effect of the modulation (see Figure \ref{fig:main4} e). We find that the contrast reduces with increasing mean energy. A linear fit to the contrast leads to a slope of $dC/dE_\text{exp.} = -0.075(13)/\hbar \omega_\text{r} $.
We compare this value to the slope of $-0.048(3) /\hbar \omega_\text{r}$ that we extract from a simulation using thermal states as described in Ref. \cite{Rakshit2017} (see Figure \ref{fig:main4} f).
While our finite, non-interacting system is not expected to thermalize after the modulation, the number of excited states that might contribute to the density matrix at a given excitation energy is very large. The dimension of the Hilbert space of excitations from the $N=6$ particle ground state to the next few higher shells is already on the order of a few hundred thousands. Trap imperfections like anharmonicity and anisotropy lead to coupling between the different degrees of freedom. Together with small potential drifts this motivates our comparison of the measured final state to a thermal mixture. In addition to deviations from a thermal state, the discrepancies we find may be due to the systematic uncertainties in determining the total kinetic energy or due to additional excitations in the axial direction. The thermalization dynamics that may occur in the presence of interactions are an exciting topic for future studies.

In conclusion, we find that the finite temperature of our experiments is one factor that contributes to the reduced contrast of the measured Pauli crystals. The ability to melt the Pauli crystal clearly shows that the observed correlations originate from the fermionic nature of our initial state. Neither fidelity nor resolution of our imaging technique depend on the initial state energy.

We have observed that Pauli's principle leads to the formation of striking geometric configurations of fermions confined in a trap, even in the absence of any interactions. The structure is not apparent in the density distribution directly but only reveals itself in correlations between relative positions or momenta. Each single experimental realization still fluctuates and can deviate significantly from the most probable configuration.

Many interacting mesoscopic systems, like ions \cite{Wineland1987}, Rydberg atoms \cite{Schauss2015} or dipolar gases \cite{Kadau2016} show self-ordering and crystalline structures akin to what is observed here. While this motivates the term Pauli \textit{crystal}, we stress that in our case translational symmetry is not broken and no long range order is present. The ground state is a coherent superposition of all possible configurations and the rotational symmetry is only broken through the actual measurement.

We generically expect this kind of order driven by quantum statistics to be present in few-fermion systems of fixed particle number. Our simulations show that similar structures appear, for example, also in box potentials. While the exact trapping potential is not important, the order will be pronounced as long as the interparticle spacing is not much smaller than the size of the system. As the system size is increased, we expect the structures to decrease in contrast until they vanish for a homogeneous, infinite Fermi gas.

Our measurements demonstrate that the correlation environment of individual particles can now be accessed in continuum systems. This unique capability will be extremely useful for future studies of correlations in strongly interacting systems. Our single atom imaging scheme can be made sensitive to a second spin state \cite{Bergschneider2018}, which may directly reveal pairing correlations near the few-body precursor of a phase transition that we have recently observed \cite{Bayha2020}.
Scaling up the system size will enable us to shed further light on many open questions concerning two-dimensional Fermi gases like the nature of its normal phase \cite{Murthy2018} or to study the emergence of Cooper pairing \cite{Altman2004}. The detection of momentum correlations represents one milestone on the path towards understanding many of these complex fermionic many-body systems.


\paragraph*{Acknowledgements}
This work has been supported by the Heidelberg Center for Quantum Dynamics,  the DFG Collaborative Research Centre SFB 1225 (ISOQUANT) and the European Union’s Horizon 2020 research and innovation program under grant agreement No.~817482 PASQuanS and grant No.~725636. K.S. acknowledges support by the Landesgraduiertenf\"orderung Baden-W\"urttemberg. P.M.P acknowledges funding from the Daimler and Benz Foundation.

\paragraph*{Author Contributions}
L.B.\ and M.H.\ contributed equally to this work. L.B.,\ M.H.,\ and K.S.\ performed the measurements and analyzed the data. C.H.\ and M.H.\ performed the numerical calculations. P.M.P.\ and S.J.\ supervised the project. All authors contributed to the discussion of the results and the writing of the manuscript.



%

\setcounter{figure}{0}
\renewcommand{\figurename}{Extended Data Figure}
\cleardoublepage
\newpage
\section*{Supplemental Material}
\paragraph*{\textbf{Experimental sequence}}

The experimental sequence starts by transferring a Fermi gas of $^6$Li atoms from a magneto-optical trap (MOT) into a red-detuned crossed-beam optical dipole trap (ODT). Here, we make use of radio frequency pulse sequences to prepare a balanced mixture of the two electronic hyperfine states $\ket{1}$ and $\ket{3}$ of the ground state of $^6$Li. We label the hyperfine states according to their energy from lowest ($\ket{1}$) to highest ($\ket{6}$).

After a first evaporate cooling stage in the ODT we transfer around 1000 atoms into a tightly focused optical tweezer (OT). In the OT quantum degeneracy is reached by spilling to around 20 atoms close to the ground state of the approximately harmonic confinement by the procedure described in Ref.\ \cite{Serwane2011a}. Subsequently, we begin the crossover to a quasi two-dimensional system. This is achieved with an adiabatic transfer of the atoms from the effective confinement of the OT alone with $\omega_\text{r}: \omega_\text{z} \approx 5:1$ to $\omega_\text{r}: \omega_\text{z} \approx 1:7$ in a combined potential of OT and a single layer of a  one dimensional optical lattice. To this end, we lower the radial trap frequency $\omega_\text{r}$ of the OT from approximately $2\pi\times20\,\text{kHz}$ to $2\pi\times 983(5)\,\text{Hz}$ by reducing the aperture of the OT setup with a Spatial Light Modulator in $20\,\text{ms}$. This changes the waist of the OT from $\approx 1 \, \mu \text{m}$ to $\approx 5 \, \mu \text{m}$. The axial confinement is solely defined by the optical lattice with $\omega_\text{z}= 2\pi\times 6560(6)\,\text{Hz}$. The anisotropy of the combined potential in radial direction  is on the order of $2\,\%$. The finite size of the Gaussian beam of the OT leads to a finite anharmonicity and to a transition frequency of the lowest to the second shell that is $10\,\%$ larger than the transition from the second to the fourth shell.

In the combined trap, we prepare closed shell configurations of the quasi-2D harmonic oscillator, by applying a magnetic gradient of approximately $70\, \text{G/cm}$ in axial direction and reducing the power of the OT such that only the lowest two ($N=3$) or three ($N=6$) levels remain bound. The spilling procedure is performed at a magnetic offset field of $B=300\text{G}$, where the interaction energy is sufficiently small that one recovers the non-interacting shell structure. We find that a finite negative scattering length improves the fidelity of the spilling process compared to a completely non-interacting sample. After preparation, we increase the OT power back up until we recover the trap frequencies and aspect ratio discussed above. In the last step we ramp the magnetic field to the zero crossing of the scattering length at $B=568\,\text{G}$.

The measurements of the Pauli crystal are performed by a time-of-flight expansion of the initial state in the single layer of the lattice that is also responsible for the axial confinement. With a radial confinement of $\omega_\text{R} = 2 \pi \times 20.7(5) \,\text{Hz} $ this corresponds to a magnification of the initial state wave function by a factor of $\omega_\text{r}/\omega_\text{R}\approx50$ after expanding for a quarter trap period. To resolve single atoms we collect the fluorescence signal that the atoms emit in free space when excited by two counter propagating resonant beams for $20\,\mu\text{s}$ \cite{Bergschneider2018}.

\paragraph*{\textbf{Theoretical Background}}
The single particle wavefunctions of a symmetric two-dimensional harmonic oscillator are given by
\begin{equation}
    \Psi_{p,q}(x,y)=\frac{\text{e}^{-(x^2+y^2)/2}\mathcal{H}_p(x)\mathcal{H}_q(y)}{\sqrt{2^{p+q}p!q!\pi}}\text{,}
\end{equation}
where $\mathcal{H}_p(x)$ is the $p$-th Hermite polynomial and x and y are expressed in units of $l_\text{0}=\sqrt{\hbar/m \omega_\text{r}}$. The energy of the single particle states is given by $E_{pq}=\hbar\omega_\text{r}(p+q+1)$. This leads to degenerate energy levels with total excitation number $n = p+q = 0,1,2,...$, where the $n$-\text{th} energy level is $(n+1)$-fold degenerate (see Figure \ref{fig:main1} a). The non-interacting $N$-body ground state is given by the configuration where the lowest $N$ single particle energy levels up to the Fermi energy are occupied. For $N=1,3,6,10,\,...$ this results in uniquely defined closed shell configurations. The $N$-body wavefunction in real space is the Slater determinant of the occupied single particle wavefunctions $\Psi_i$:
\begin{equation}
    \Psi(r_1, ..., r_N) = \sqrt{\frac{1}{N!}}\text{det}\left[ \Psi_i (r_k) \right].
\end{equation}
The single particle wavefunctions $\Psi_{p,q}$ of the harmonic oscillator are eigenfunctions of the continuous Fourier transform. Since each term of the Slater determinant contains the same product over single particle states $\Psi_{i}$ $(i=1...N)$, the $N$-body ground state $\Psi(r_1, ..., r_N)$ is an eigenfunction of the Fourier transform as well.

\begin{figure}[hbt!]
    \centering
	\includegraphics{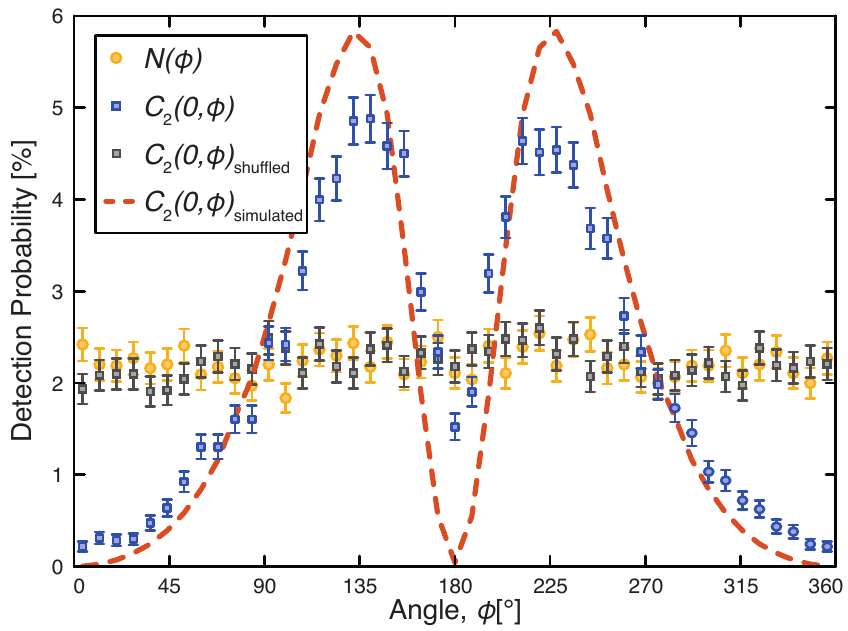}
    \caption{
    \textbf{Angle correlations for $\bm{N=3}$ particles.} The angular correlation function $C_2$ is computed after removing the center of mass motion, by fixing one particle at $\phi_1=0$ and creating a histogram of the measured angles of the other particles. $C_2(0,\phi)$ shows 2 maxima in addition to the hole around $\phi=0$ as expected for the 3-fold symmetry of the Pauli crystal (blue points). The measurement agrees well with a Monte Carlo simulation (dashed line). Both the total angular atom distribution $N(\phi)$ (yellow points) and the correlation function of shuffled data (grey points) show no structure.}
    \label{fig:som6}
\end{figure}

\begin{figure*}[hbt!]
    \centering
	\includegraphics{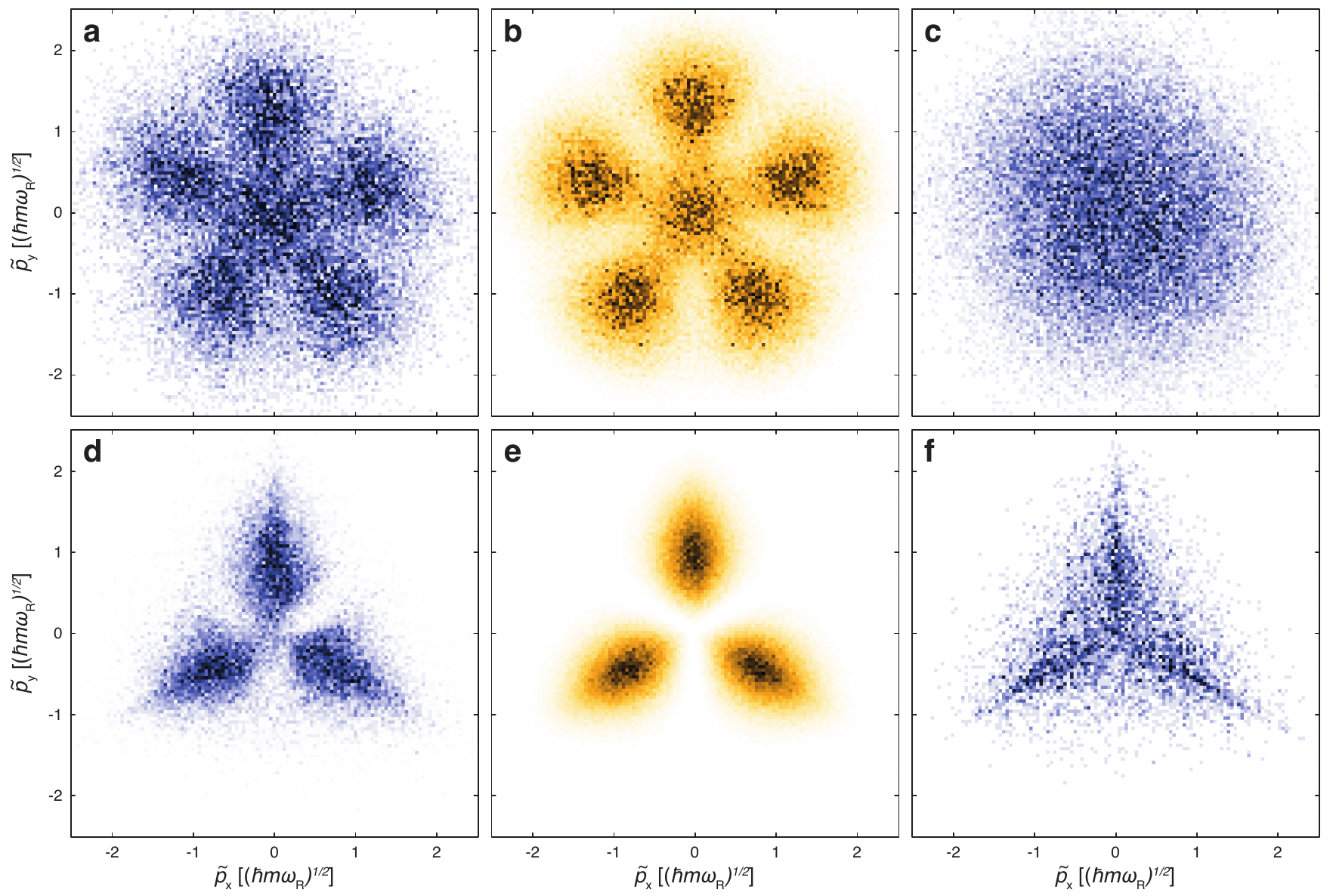}
    \caption{\textbf{Comparison to Monte Carlo simulations and shuffled data.} We compare our measured Pauli crystals (a,d) to images that we obtain by sampling from the zero temperature $N$-body wavefunction (d,e). To this end we apply a Markov chain Monte Carlo method as described in Ref. \cite{Rakshit2017}. In (c) and (f) we show the configuration densities that we obtain after shuffling the atom momenta $p_i$ between different experimental runs. We observe no bias of our evaluation method towards the target configuration for $N=6$. The residual structure that can be observed for $N=3$ is a boundary effect that appears at this small particle number due to the limited number of degrees of freedom.}
    \label{fig:som5}
\end{figure*}

\paragraph*{\textbf{Image Analysis}}
To reveal correlations between the atoms, we analyse our data in two steps. First we subtract the center of mass momentum $\bar{p}$ from each set of momenta $p_i$ $(i=1...N)$. In a second step we rotate each image separately by an angle $\alpha$ to align all images to the same symmetry axis. The angle $\alpha$ is determined by minimizing the distance of the rotated configuration $\tilde{p}_i(\alpha)$ to a chosen target configuration $p_{\text{T},i}$ as follows:
\begin{equation}
    \alpha = \text{mean}(\tilde{\phi}_i(\alpha) - \phi_{\text{T},i}).
\end{equation}
Here, $\tilde{\phi}_i(\alpha)$ and $\phi_{\text{T},i}$ are the angles in the center of mass coordinate system of atom $i$ for the rotated and target configuration respectively. The index $i$ runs clockwise over all the atoms in the outer shell of the given Pauli crystal (i.e. $i=1...3$ for $N=3$ and $i=1...5$ for $N=6$). To check for a possible bias of the configuration density, we compare to data where the atom momenta have been shuffled between different experimental runs (see Extended Data Figure \ref{fig:som5}). While a small bias is present for $N=3$ particles due to the small number of degrees of freedom of this system, we find that this is not the case for any larger particle number. In addition, the angle density correlations function $C_2$ reveals the same $3$-fold and $5$-fold symmetries of the $N=3$ and $N=6$ Pauli crystals respectively (see Figure \ref{fig:main3} and Extended Data Figure \ref{fig:som6}).

\cleardoublepage
\begin{figure}[hbt!]
    \centering
	\includegraphics{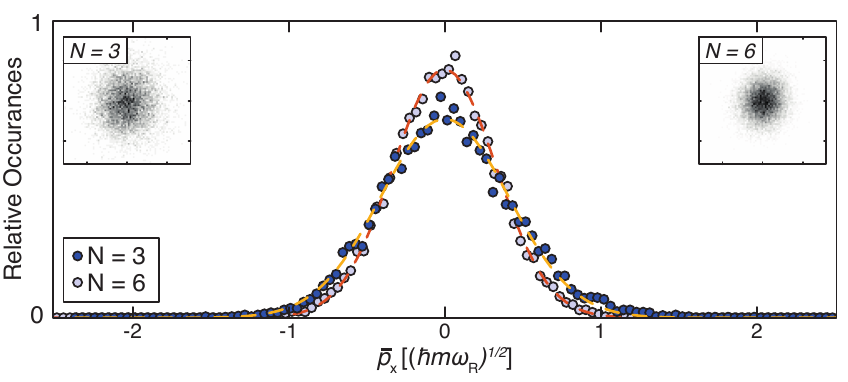}
    \caption{ \textbf{Center of mass motion.} Our few-body system shows fluctuations of the center of mass coordinate around its expectation value. By increasing the particle number from $N=3$ (dark blue) to $N=6$ (light blue), we double the mass of the center of mass system. This leads to a reduction of the fluctuations of the center of mass momentum by a factor of $1.40(1) \approx \sqrt{2}$ as expected (insets).
    The absolute width of the distributions are given by $\sigma_{N=3}=0.44(1)\,p_\text{0}$ and $\sigma_{N=6}=0.31(1)\,p_\text{0}$. The expected values for the ground state wavefunction are $\sigma_{N=3}=1/\sqrt{2\cdot N}=0.41\,p_\text{0}$ and $\sigma_{N=6}=0.29\,p_\text{0}$. This shows that quantum fluctuations dominate the observed distributions, which are broadened slightly by classical noise, for example through fluctuations in the potential that is used for the TOF expansion. The dashed lines are Gaussian fits to the data.}
    \label{fig:som1}
\end{figure}
\newpage


\begin{figure*}[hbt!]
    \centering
	\includegraphics{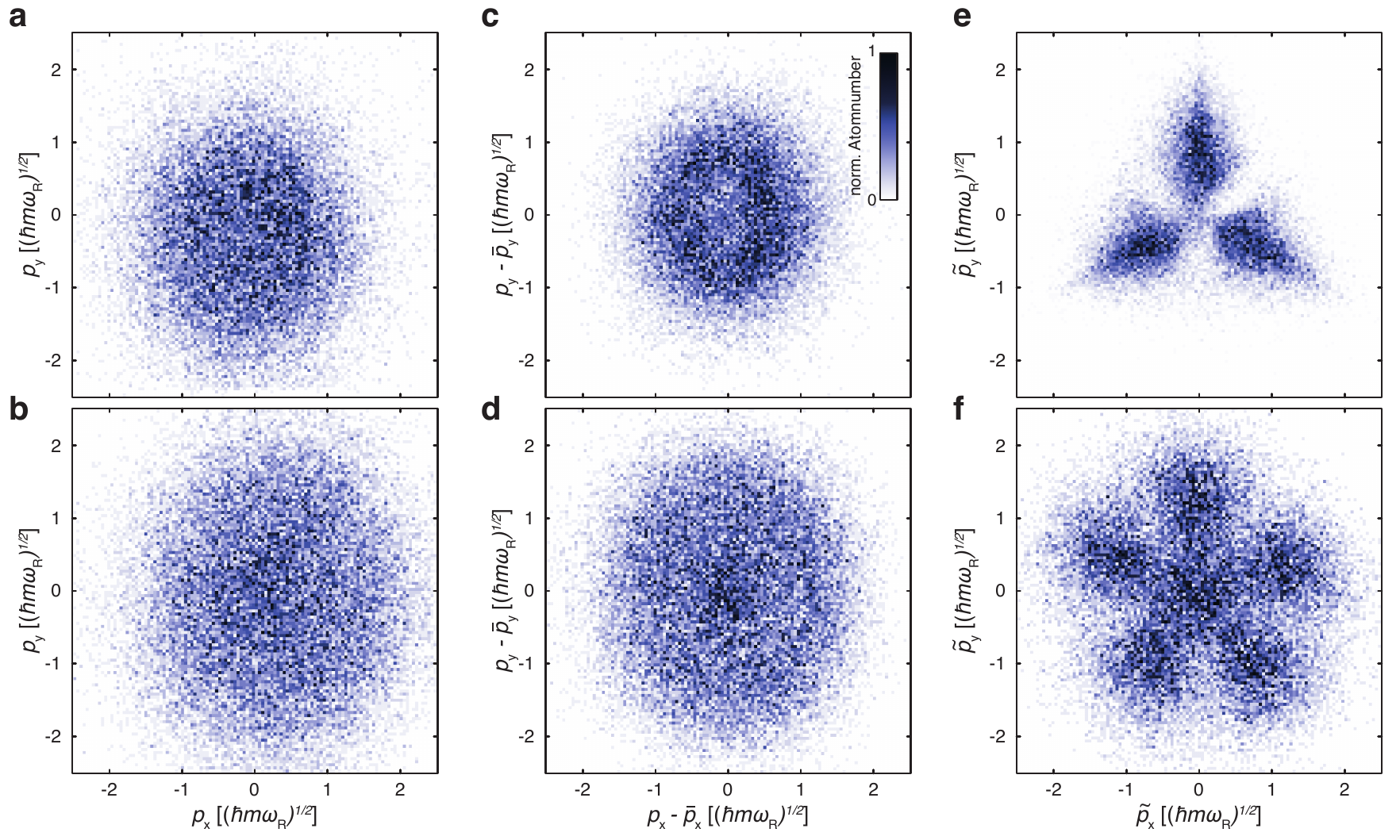}
    \caption{ \textbf{Analysis procedure.} By plotting all measured particle momenta into a single histogram, we obtain the one particle momentum density for $N=3$ (a) and $N=6$ (b). Their deviations from Gaussian functions become more apparent once we subtract the center of mass momentum (c,d). Strong correlations between the respective particles as result of the Pauli principle only reveal themselves after defining a common symmetry axis. This leads to the configuration probability density (e,f).}
    \label{fig:som2}
\end{figure*}

\end{document}